\documentclass[twocolumn,twoside,slac_two]{revtex4}
\usepackage{graphicx}
\usepackage{fancyhdr}
\def \bs         {B^0_s}
\def \bsbar      {{\bar B}^0_s}

\def \bzero      {B^0}
\def \bzerobar   {{\bar B}^0}
\def \bzeroToPsiKShort {\bzero \rightarrow J/\psi K^0_s}
\def \bsToPsiPhi {\bs\rightarrow J/\psi\phi}

\begin{document}

\title{$CP$ Violation in $\bs$ Mesons}

%

\author{J. Boudreau}
\affiliation{
Department of Physics and Astronomy, 
University of Pittsburgh, Pittsburgh PA 15260 \\
For the CDF and D0 Collaborations. 
}

\begin{abstract}

The $\bs$ meson is a bound state of b and s type quarks.  A $CP$ violation parameter, $\beta_s$, 
of that system is the analogue of the parameter $\beta$ measured precisely at the $B$ factories
in $\bzero$ decays.  The standard model predicts, robustly and precisely,  a value of $\beta_s$
which is very close to zero.  The CDF and D0 experiments now have about 2000 fully
reconstructed and flavor-tagged $\bsToPsiPhi$ decays each, with which they set
new experimental bounds on $\beta_s$. A combination of results 
from CDF and D0 is consistent with the standard model at only the 2.2 $\sigma$ level.  
If the discrepancy is not a statistical fluctuation, it would indicate new 
sources of $CP$ violation. 


\end{abstract}

\maketitle

\thispagestyle{fancy}


\section{Introduction}

For many years, the neutral kaon system was the only place in which
the violation of $CP$ symmetry was observed\cite{ref:croninfitch}.  The last decade has
witnessed an intensive effort to record and interpret as many cases of
$CP$ violation in neutral and charged $B$ mesons as possible.  The CDF and
D0 experiments now, for the first time, are able to extend the search
for $CP$ violation to the neutral $\bs$ meson.  This system combines
observable fast particle-antiparticle oscillations familiar from the 
$\bzero$ system, with an observable separation into distinct lifetime 
states best known from the neutral kaon system. $CP$ violation in the $\bs$
system, the subject of this paper, contains some elements resembling
$CP$ violation in the kaon system, and others that resemble $CP$ violation
in the $\bzero$ system. 

\begin{figure}[h]
\centering
\includegraphics[width=80mm]{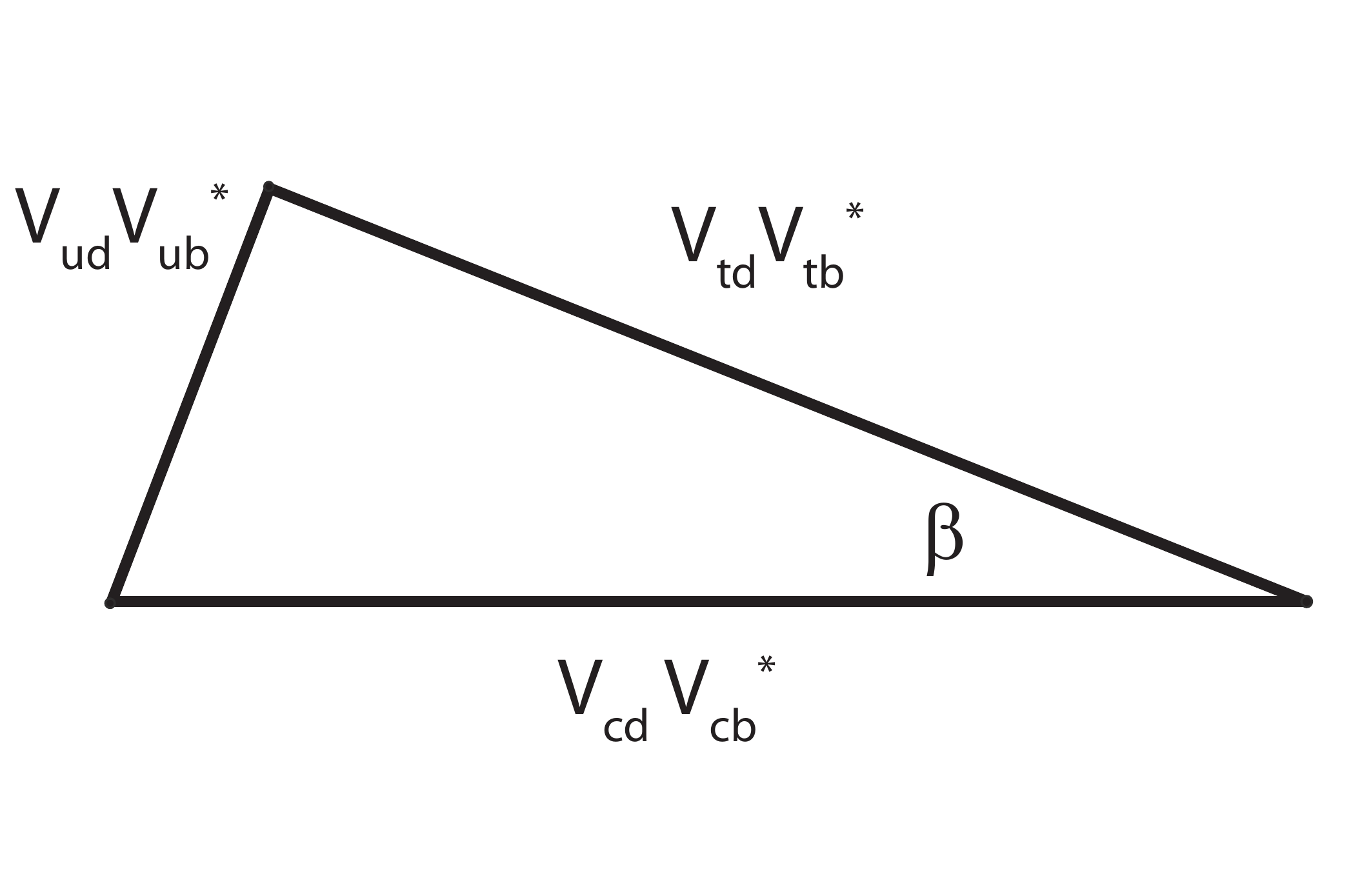}
\caption{The usual (bd) unitarity triangle, showing the angle $\beta$, measured
precisely in decays like $B^0\rightarrow J/\psi K^0_s$ at the B-Factories. All sides of this triangle are ${\cal O} (\lambda^3)$} \label{fig:unitarity1}
\end{figure}

\begin{figure}[h]
\centering
\includegraphics[width=80mm]{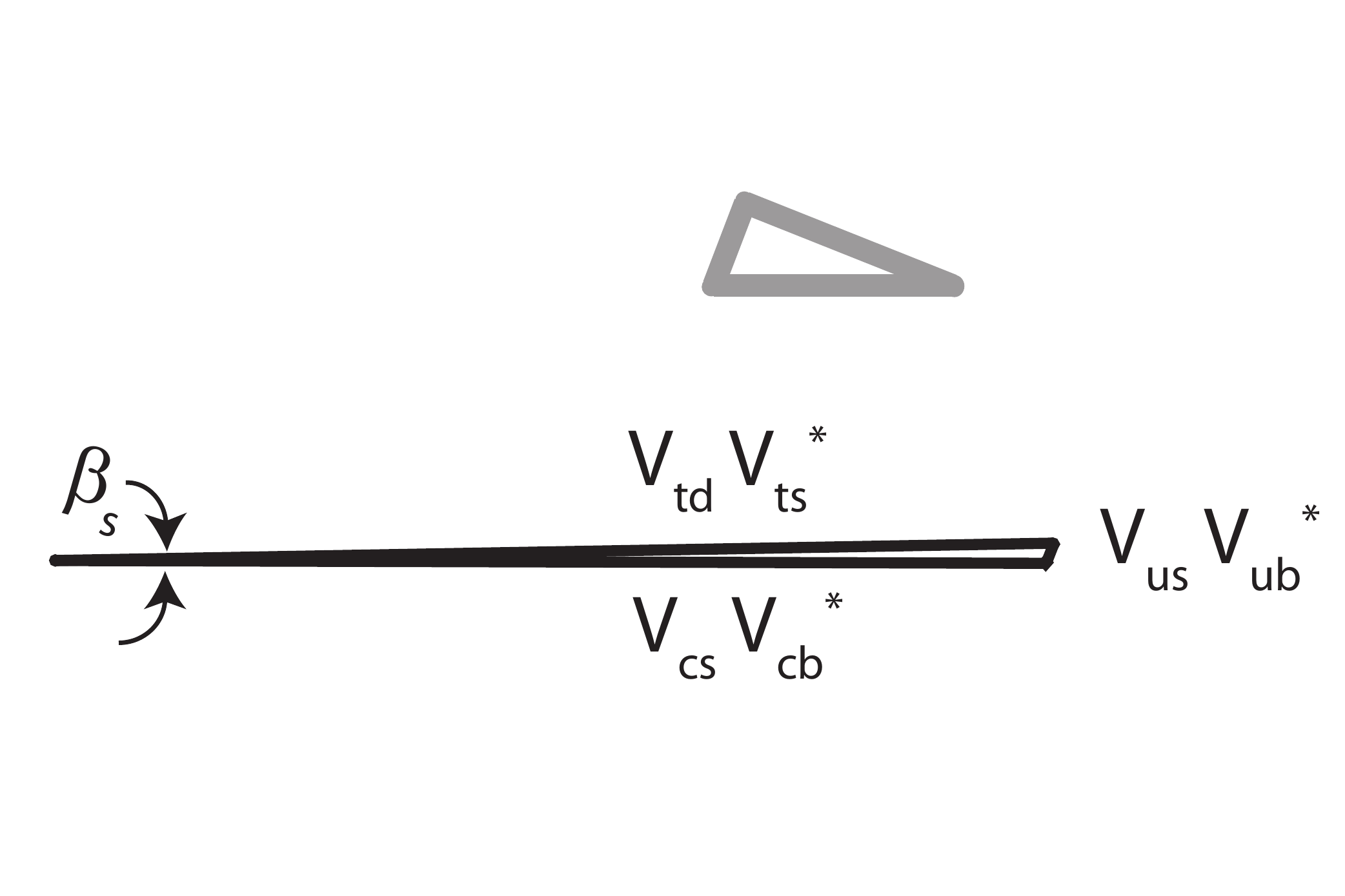}
\caption{The ``squashed'' (bs) unitarity triangle showing $\beta_s$,
the angle at the most acute vertex of the triangle.   This triangle 
has two sides of length {\cal O}($\lambda^2$) and a third side of
length {\cal O}($\lambda^4$). For comparison the (bd) unitarity triangle is drawn, to scale, in light gray. } 
\label{fig:unitarity2}
\end{figure}

One of the manifestations of the large $CP$ violation in the neutral
$\bzero$ is the $CP$ asymmetry in certain decays such as
$\bzeroToPsiKShort$.  This $CP$ asymmetry is characterized by the angle
\begin{equation}
\label{beta}
\beta = {\rm arg}\left( - \frac{V_{cd}V_{cb}^{*}}{V_{td}V_{tb}^{*}} \right )
\end{equation}
of the unitarity triangle, shown in
Fig.~\ref{fig:unitarity1}. The groundbreaking measurements\cite{ref:Babar,ref:Belle} of the angle
$\beta$ stand out from the dozen or so other measurements of $CP$
violation\cite{ref:pdg}, because the measurement is experimentally
clean and the prediction is largely free from theoretical
uncertainties. In the neutral $\bs$ system, the same comments apply to
the decay $\bsToPsiPhi$, where one can measure the quantity
\begin{equation}
\label{betas}
\beta_s = {\rm arg} \left( - \frac{V_{ts}V_{tb}^{*}}{V_{cs}V_{cb}^{*}}
\right ),
\end{equation}
an angle of the ``squashed'' (bs) unitarity triangle (Fig.~\ref{fig:unitarity2}), whose standard
model value is $\lambda^2\eta=$ 0.019 $\pm$ 0.001, where 
$\lambda$=0.2257$^{+0.0009}_{-0.0010}$ and $\eta=$0.349$^{+0.015}_{-0.017}$
are parameters of the CKM matrix\cite{ref:wolfen}. It is measurable\cite{ref:dighe2} in the decay 
$\bsToPsiPhi$, through the interference of mixing and decay. The measurement is sensitive 
to new physics, particularly if it affects the $t\rightarrow s$ transition.  
One such scenario is discussed in reference \cite{ref:hou}.

\section{The decay $\bsToPsiPhi$}

It is useful to think of the $CP$-mixed $J/\psi \phi$
as three distinct final states, characterized either by the orbital angular
momentum of the two vector mesons $J/\psi$ and $\phi$, or by their relative 
polarization  $\left \{ 0, \parallel, \perp \right \}$, where the first 
symbol indicates longitudinal polarization vectors, the second indicates transverse polarization vectors
which are mutually parallel, and the third indicates transverse polarization vectors
which are mutually perpendicular. We designate the three states as ${\cal P}_0$, ${\cal P}_\parallel$ and
${\cal P}_\perp$ the first two being $CP$-even and the third $CP$-odd.
It is also useful to think of the $\bs$ as two distinct initial states, 
the long-lived ``heavy'' and short-lived ``light'' mesons:
\begin{displaymath}
|B_s^H\rangle = p\,|B_s^0\rangle - q\,|\bar B_s^0\rangle, 
\hspace*{0.9cm} 
|B_s^L\rangle = p\,|B_s^0\rangle + q\,|\bar B_s^0\rangle. 
\end{displaymath}
$CP$ violation in this system presents itself in two ways. If $[H,CP]\ne 0$ then
the long- and short-lived mass eigenstates are {\it not} $CP$  eigenstates and
may decay to both $CP$ even or $CP$ odd final states. This is reminiscent of the
neutral kaon system.  Additionally, the expectation value  $\langle CP \rangle$
from an initially pure $\bs$ or $\bsbar$ evolves with time:  $d\langle CP \rangle /dt \ne 0$.
The time evolution is an oscillation with $\bs$ mixing frequency of 
$\Delta m_s$=17.77 $\pm$ 0.10 $\pm$ 0.07 ps$^{-1}$\cite{ref:cdfbs}.  The time-dependent $CP$ expectation 
reflects itself in a time-dependent polarization of the two vector mesons, and finally in a time variation of the 
angular distributions of their decay products.  This is similar to the situation in 
the $\bzero$ system, particularly in the $B\rightarrow VV$ decay  $\bzero\rightarrow J/\psi K^{0*}$. 
The CDF and D0 analyses, which fit the differential rate of $\bsToPsiPhi \rightarrow \mu^+\mu^-K^+K^-$, 
are simultaneously sensitive to both effects.

The time-dependent rates for initially pure $\bs$ and $\bsbar$  are
\begin{eqnarray} 
P_{B}({\hat n},\psi, t) = \frac{9}{16\pi} |{\bf A}(\psi,t)\times \hat{n}|^2  \nonumber \\
P_{\bar{B}}({\hat n}, \psi, t) = \frac{9}{16\pi} |{\bf {\bar A}}(\psi,t)\times \hat{n}|^2
\label{eqn:master}
\end{eqnarray}
\cite{ref:dighe} where ${\hat n}$ is the direction of the $\mu^+$ in the rest frame of the $J/\psi$, $\psi$ is the 
helicity angle in the $\phi$ decay,  and ${\bf A}(\psi,t)$ and ${\bf {\bar A}}(\psi,t)$ are complex vectors\footnote{Complex conjugation is implied in the square magnitude.}, 
described below, reflecting the time-dependent  polarization.  
A coordinate system is needed to express the vectors ${\hat n}$, ${\bf A}$, and ${\bf {\bar A}}$: two
common choices are the {\it transversity basis}, and the {\it helicity basis}.   
The transversity basis has the  $x$-axis along the $\phi$ direction in the rest frame of the $B$, 
the $y$-axis lying in  the decay plane of the $\phi$ such that $P_{y}(K^+)>$0, and ${\hat z} = {\hat x} \times {\hat y}$.  The
helicity basis is a cyclic permutation of the axes in the transversity basis:  ${\hat x_T} = {\hat z_H}$,
etc. CDF and D0 employ the transversity basis, in which 
${\hat n}= (\sin{\theta_T}\cos{\phi_T}, \sin{\theta_T}\sin{\phi_T}, \cos{\theta_T})$ and
\begin{eqnarray*}
{\bf A}(\psi,t)=({\cal A}_0(t)\cos{\psi}, -\frac{{\cal A}_\parallel(t)\sin{\psi}}{\sqrt{2}}, i\frac{{\cal A}_\perp(t)\sin{\psi}}{\sqrt{2}})  \nonumber \\
{\bf{\bar {A}}}(\psi,t)=({\bar {\cal A}}_0(t)\cos{\psi}, -\frac{{\bar {\cal A}}_\parallel(t)\sin{\psi}}{\sqrt{2}}, i\frac{{\bar {\cal A}}_\perp(t)\sin{\psi}}{\sqrt{2}})
\label{eqn:fixedAngle}
\end{eqnarray*}
where the time dependence is contained in the term
\begin{eqnarray}
{\cal A}_i(t) = A_i(0)\left[
E_+(t) \pm e^{2i\beta_s} E_-(t)\right],   \nonumber \\
{\bar {\cal A}_i}(t) = A_i(0) \left[ \pm E_+(t) + e^{-2i\beta_s} E_-(t),
\right]  
\label{eqn:finalAmp}
\end{eqnarray}
where:
\begin{displaymath}
E_{\pm}(t) \equiv  \frac{e^{-t/2{\bar \tau_s}}}{2}\left[e^{+\left(\frac{-\Delta\Gamma_s}{4} + i\frac{\Delta m}{2}\right)t} \pm e^{-\left(\frac{-\Delta\Gamma_s}{4} + i\frac{\Delta m}{2}\right)t}\right].
\label{eqn:functionDef}
\end{displaymath}
and ${\bar \tau}_s$ is the $B^0_s$ mean lifetime. Two strategies can be pursued.  One can measure the differential rates given in Eq.~\ref{eqn:master}, without
attempting to distinguish $\bs$ from $\bsbar$, effectively summing the rates over both species.  Or, one can 
try to measure the differential rates for $\bs$ and $\bsbar$ separately, using a technique called flavor
tagging. 

Eq.~\ref{eqn:master} contains a lot of hidden richness, including a measurable width difference between
two mass eigenstates, $CP$ asymmetries that are measurable in a flavor-tagged analysis, with simultaneous sensitivity
to {\it both} $\sin{2\beta_s}$ and $\cos{2\beta_s}$. Even without flavor tagging, a residual sensitivity to $CP$ 
violation is present.  This arises partly due to the ability to detect the decay of the short-lived, nominally $CP$-even
mass eigenstate to the $CP$-odd polarization states (and vice-versa), which was historically the basis of the first observation 
of $CP$ violation.  And partly it arises from the interference between the even and odd polarization states of the vector mesons, 
which are, in fact ``intermediate'', not ``final'' states. Even in the case that $[H,CP]=0$, the differential rates in Eq.~\ref{eqn:master} have a sensitivity to $\bs$-$\bsbar$ oscillations,
though neither experiment can exploit it with present statistics.

\section{Detector Effects and their Impact}
\label{section:methods}

In each event one measures, in addition to the proper decay time $t$, the kinematic quantities 
$\cos{\psi}$, $\cos{\theta_T}$, and $\phi_T$.   Three significant detector effects alter the theoretical 
model (Eq.~\ref{eqn:master}).  First, measurement uncertainty smears significantly the oscillatory time dependence of the rates.  
Second, the flavor tagging algorithms in use at the Tevatron give very limited discrimination between 
$\bs$ and $\bsbar$  mesons.  Third, the detector acceptance alters the angular distributions in 
Eq.~\ref{eqn:master}.   The first two effects limit the measurements, 
while the third does not have a significant impact if properly accounted for. 

\begin{figure*}[t]
\centering
\includegraphics[height=70mm]{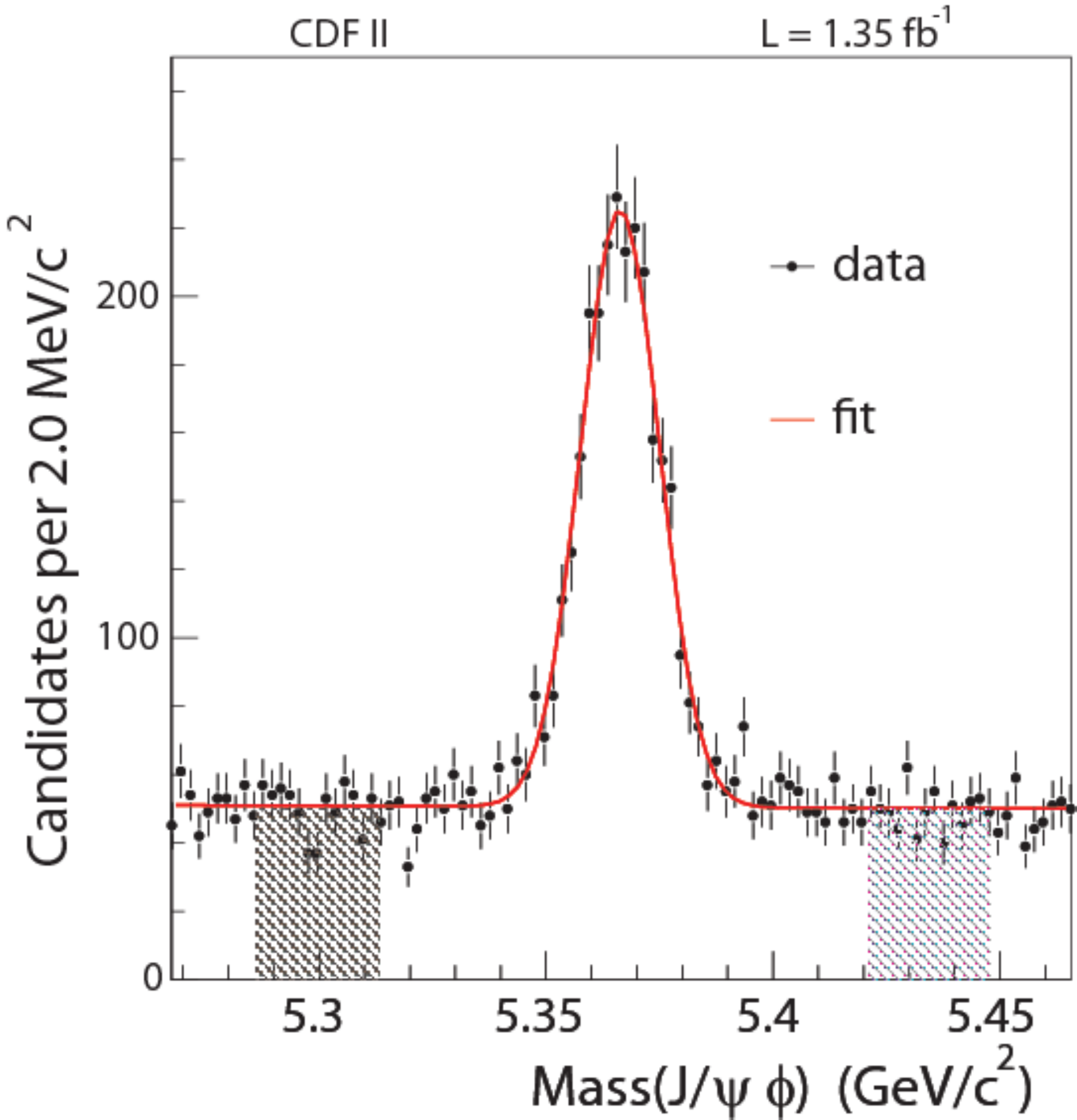}
\includegraphics[height=70mm]{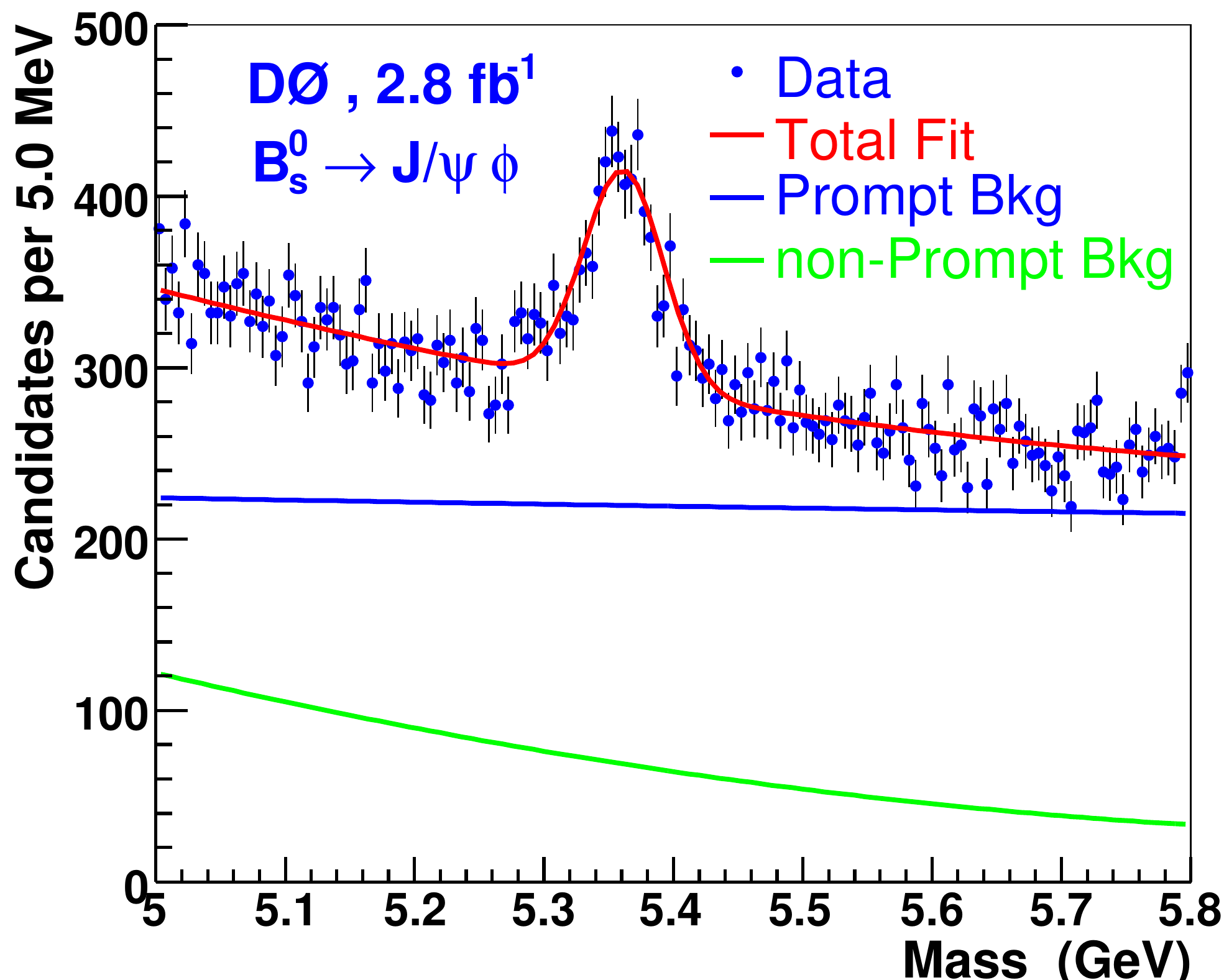}
\caption{Data samples for the CDF tagged analysis of $\bsToPsiPhi$.  The CDF experiment reports
2019$\pm$73 events while the D0 experiment reports 1967 $\pm$ 65 events.} \label{fig:dataSamples}
\end{figure*}


Flavor tagging is an essential ingredient for many studies of $B$ mixing and $CP$ violation.  Flavor
tagging endeavors to determine, from the tracks lying near to, or far from, a reconstructed meson,
the flavor of that meson ($\bzero$ or $\bzerobar$, $\bs$ or $\bsbar$) at production. 
In a decay of a neutral $B$ meson to a flavor-specific final state, the flavor at production could be different from the flavor at decay,
while, in a decay to a $CP$ eigenstate like $J/\psi \phi$, the flavor at decay is undetermined.  
Three independent flavor-tagging algorithms are currently in use at CDF and D0; these are categorized as {\it same-side} tagging or {\it opposite-side} tagging
algorithms. Same-side tagging establishes\cite{ref:tagging}, on a statistical basis, the $b$-hadron flavor through
its correlation with the charge of nearby fragmentation tracks.  Two varieties of opposite side tagging
establish the flavor of the $b$-hadron on the away side, from which one infers the flavor of the
near side $B$ meson. Opposite side lepton tagging uses
a soft lepton on the away side, while opposite side jet charge tagging 
uses  the charge of a jet on the away side.   The efficiency $\epsilon$ of any tagging algorithm 
is the fraction of the events to which it can be applied.

The tagging algorithm also estimates its uncertainty.  The tag decision being a discrete variable,
the uncertainty is quantified by the dilution $D=(R-W)/(R+W)$ where $R$ is the frequency of a right decision and $W$ is the frequency of 
a wrong decision.   The effective tagging efficiency, $\epsilon D^2$, is the figure-of-merit for a 
tagging algorithm and can be related directly\cite{ref:moserRoussarie} to the expected uncertainty in 
mixing and $CP$ asymmetry measurements\footnote{However, since the analyses we describe are not purely 
measurements of a $CP$ asymmetry, one cannot use the formula $\delta(A_{cp}) = \sqrt{2/\epsilon D^2 N}$,
developed for that purpose.}.  Despite rather different tracking and particle identification technologies, CDF
and D0 both report very similar effective tagging efficiencies, $\epsilon D^2\approx$ 4.7\% in D0
and $\epsilon D^2\approx$ 4.8\% in CDF.

Another important feature of the detector systems used in the analysis is their proper time resolution.
In a mixing or $CP$ measurement, proper time resolution further degrades the uncertainty on $CP$ asymmetries
by the factor\footnote{As before, the formula is not directly applicable in these analyses.} $\exp{(-(\Delta m_s \sigma_t)^2/2)}$. Both CDF and D0 now employ low-mass, small-radius
silicon detectors called ``Layer 00'', mounted directly on the beampipe, to achieve the best possible
resolution, under 25 $\mu$m in both experiments.  This is discussed more fully in reference\cite{ref:evansHere}.

The differential rates in Eq.~\ref{eqn:master} are sensitive to the $CP$ phase $\beta_s$, the decay width difference $\Delta\Gamma_s$,
the mean lifetime ${\bar \tau_s} = 2 / (\Gamma_H + \Gamma_L )$, and the amplitudes $A_\perp(0)$, $A_\parallel(0)$, and $A_0(0)$, with 
phases $\delta_\perp$,  $\delta_\parallel$, and zero, normalized such that $|A_0(0)|^2 + |A_\parallel(0)|^2 + |A_\perp(0)|^2$ = 1.
An expansion of Eq.~\ref{eqn:master} is convolved with an event dependent proper time resolution, adjusted for detector acceptance and 
re-normalized according to one of several schemes. In the tagged analyses, event dependent dilutions are also incorporated into the probability densities.

The simultaneous transformation $\beta_s \rightarrow \pi/2 - \beta_s$, $\Delta\Gamma_s \rightarrow -\Delta\Gamma_s$,
$\delta_\parallel \rightarrow 2\pi - \delta_\parallel$, and $\delta_\perp \rightarrow \pi - \delta_\perp$, is 
an exact symmetry of the differential rates, so, decays of $\bsToPsiPhi$ alone cannot resolve the corresponding
ambiguity.  This symmetry is an experimental headache, and has a significant impact on results from
the experiments, particularly since, with presently available statistics, the two solutions are not well separated.    
Untagged analyses possess an even higher degree of symmetry, since the simultaneous 
transformation $\delta_\perp\rightarrow \pi + \delta_\perp$, $\beta_s \rightarrow -\beta_s$ is also an exact
symmetry.  

\begin{figure*}[t]
\centering
\includegraphics[height=0.4\linewidth]{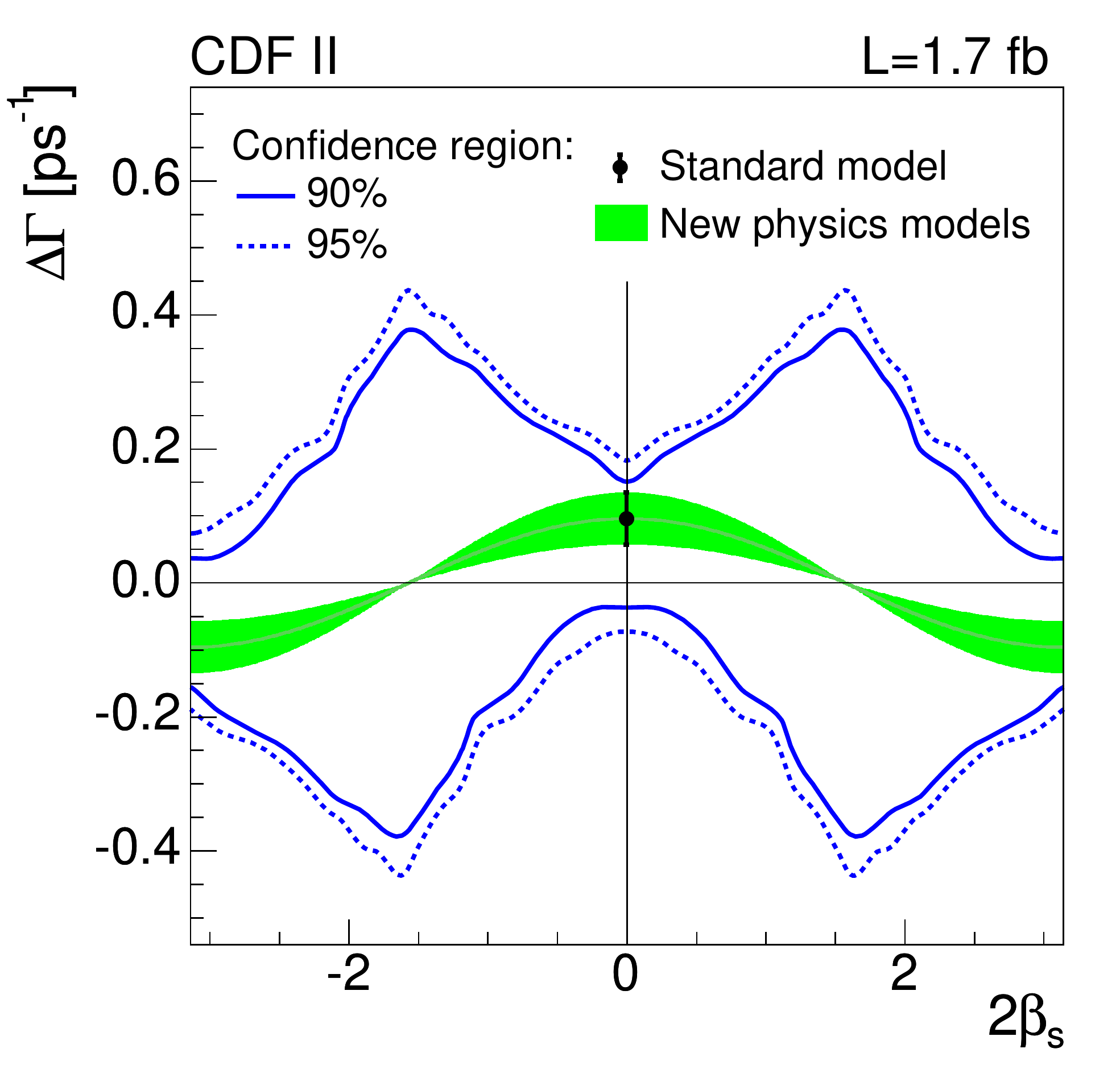}
\includegraphics[height=0.4\linewidth]{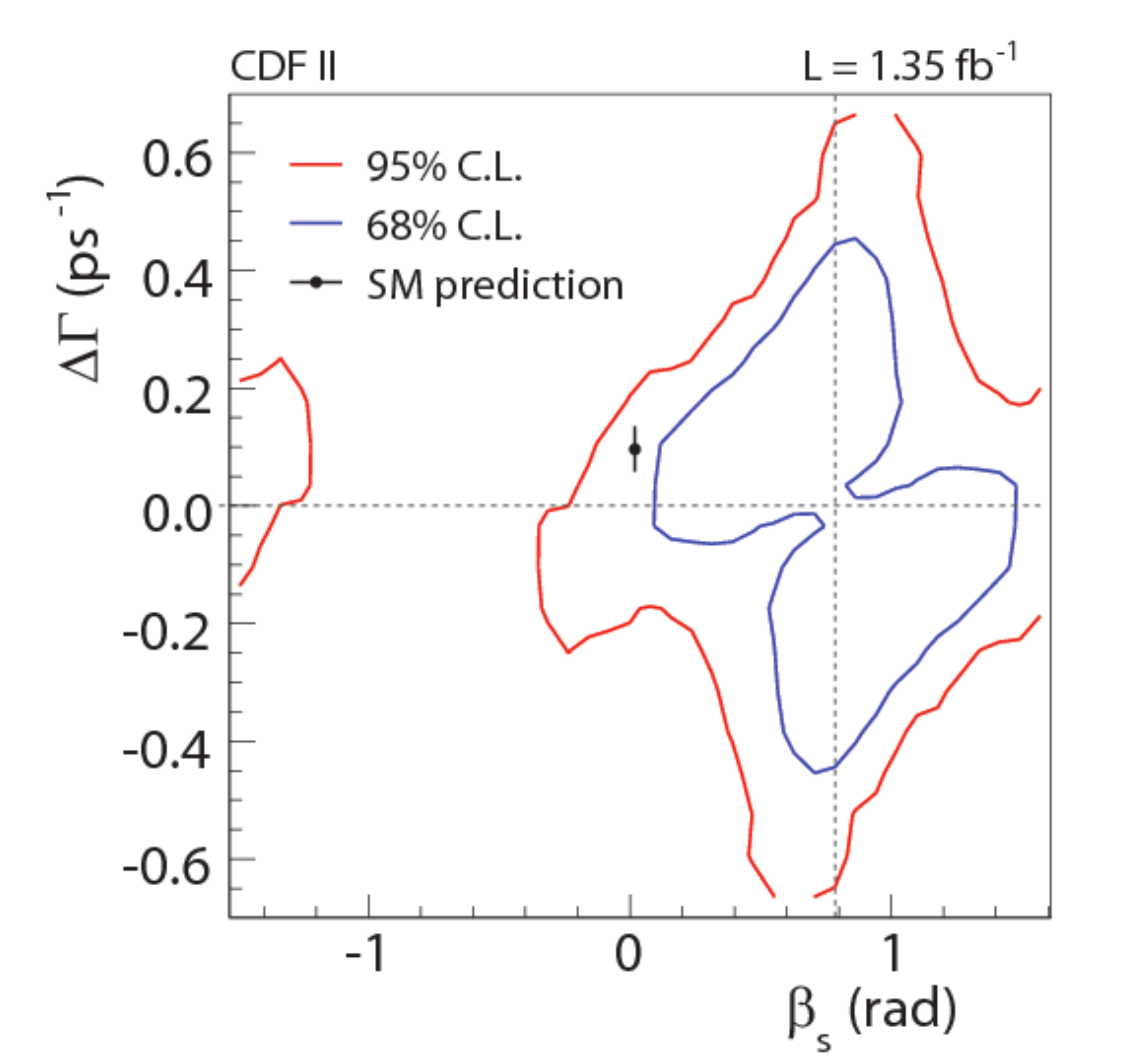}
\caption{Confidence regions in the space of parameters $\Delta\Gamma_s$ and $\beta_s$, from 1.7~fb$^{-1}$ of untagged data (left) and 1.35~fb$^{-1}$ of flavor tagged data (right). The green band corresponds to new physics models, as described in the text.}
\label{fig:cdfAnal}
\end{figure*}

\section{Results}

CDF performs an analysis without flavor tagging on a 1.7 fb$^{-1}$ sample of data\cite{ref:cdfUntagged}, and with flavor
tagging\cite{ref:cdfTagged} on a 1.35 fb$^{-1}$ sample.  The former is used for SM fits ($\beta_s = 0$) and $CP$ 
fits ($\beta_s \ne 0$), while the latter is used only for $CP$ fits.  D0 performs both SM and 
$CP$ fits to a 2.8 fb$^{-1}$ sample of tagged data\cite{ref:d0}.  Mass distributions of the signals from the two experiments 
are shown in Fig.~\ref{fig:dataSamples}. In the CDF untagged analysis, the SM
fit obtains the results shown in Table~\ref{table:standardModelFits}.  This set includes an important
measurement of $\Delta\Gamma_s$, as well as an interesting measurement ${\bar \tau}_s$,  
consistent with the HQET expectation\cite{ref:lifetimes} that ${\bar \tau}_s$ =(1.00$\pm$0.01)$\cdot \tau_0$
where $\tau_0$=1.530$\pm$ 0.009 ps is the world average $B^0$ lifetime. The amplitudes are
consistent with those measured\cite{ref:belleStrong,ref:babarStrong} in the related decay $B^0\rightarrow J/\psi K^{0*}$.
Point estimates are not obtained for  the strong phases, since the  
measurement is insensitive to $\delta_\perp$ for $\beta_s=0$ while for 
$\delta_\parallel$ the likelihood is nonparabolic (a result of the symmetries referred 
to in section~\ref{section:methods}). $CP$ fits in the untagged analysis, do not yield 
point estimates for any of the physics parameters.  They do give, however, a Feldman-Cousins confidence region\cite{ref:FeldmanCousins} shown in  Fig.~\ref{fig:cdfAnal} (left).  One can see that 
the bounds agree with the standard model (the $p$-value is 22\%, or 1.2 
Gaussian standard deviations).  Also shown in Fig.~\ref{fig:cdfAnal} (left) is 
the ``new physics'' expectation, based on the theoretical value of the decay matrix element  $2|\Gamma^s_{12}| = 0.096 \pm 0.039$ ps $^{-1}$\cite{ref:lenzNierste}, 
plus the assumption of mixing-induced $CP$ violation.  The confidence region
is in good agreement with this assumption, and cannot rule out any value of the 
$CP$ phase. The fourfold symmetry of the confidence 
region is apparent. In the CDF tagged analysis, shown in Fig.~\ref{fig:cdfAnal} (right), this symmetry is broken 
quite strongly, and about half of the parameter space for $\beta_s$ is ruled out. The 
$p$-value for the standard model is 15\%.  Statistical uncertainty dominates the measurement.  
A frequentist incorporation of systematic uncertainties\cite{ref:punzi} is included in the contour.

\begin{table}[h]
\begin{center}
\caption{Standard model fits in the CDF untagged analysis.}
\begin{tabular}{|l|c|}
\hline \textbf{Parameter} & \textbf{CDF Measurement (untagged)}                               \\
\hline  ${\bar \tau}_s = 2 / (\Gamma_H + \Gamma_L )$              &1.52  $\pm$ 0.04$\pm$0.02 ps\\
$\Delta\Gamma_s = \Gamma_H-\Gamma_L $              &0.076 $^{+0.059}_{-0.063}\pm$0.006 ps$^{-1}$\\
$|A_0|^2$                                            &0.531 $\pm$0.020 $\pm$0.007\\
$|A_\perp(0)|^2$                                        &0.239 $\pm$0.029 $\pm$0.011\\
$|A_\parallel(0)|^2$                                    &0.230 $\pm$0.026 $\pm$0.009\\
\hline
\end{tabular}
\label{table:standardModelFits}
\end{center}
\end{table}

\begin{figure}[h]
\centering
\includegraphics[width=\linewidth]{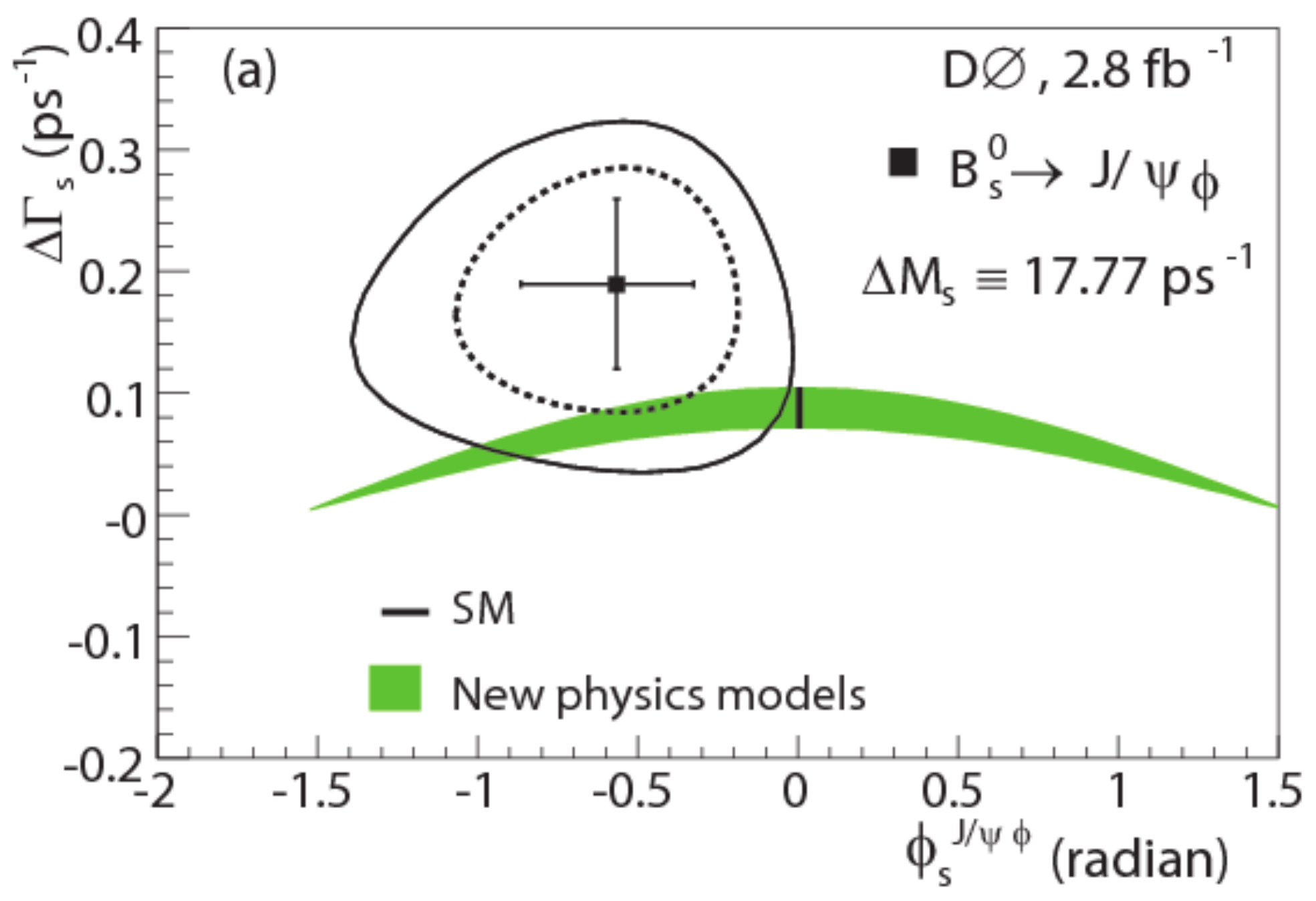}
\includegraphics[width=0.45\linewidth]{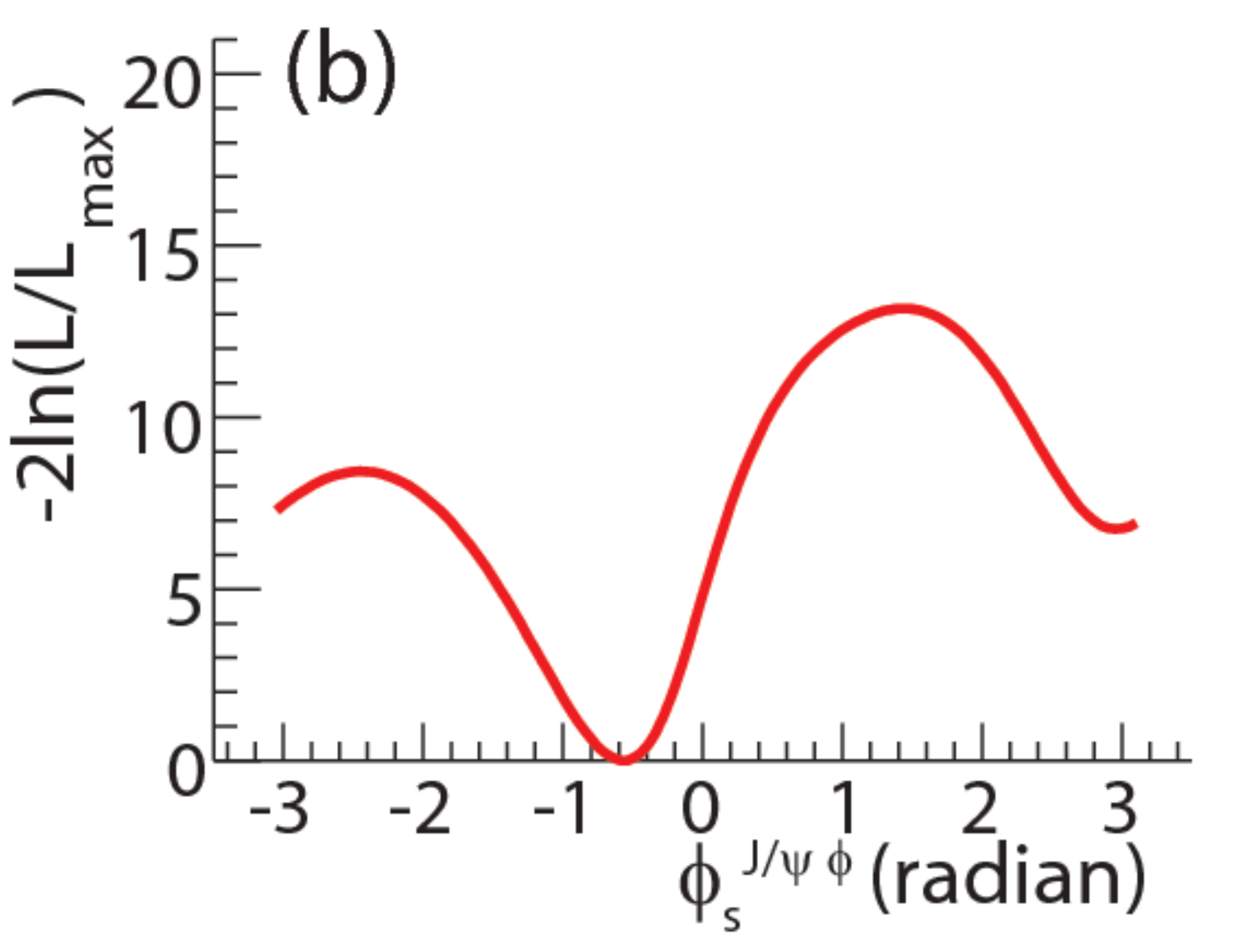}
\includegraphics[width=0.45\linewidth]{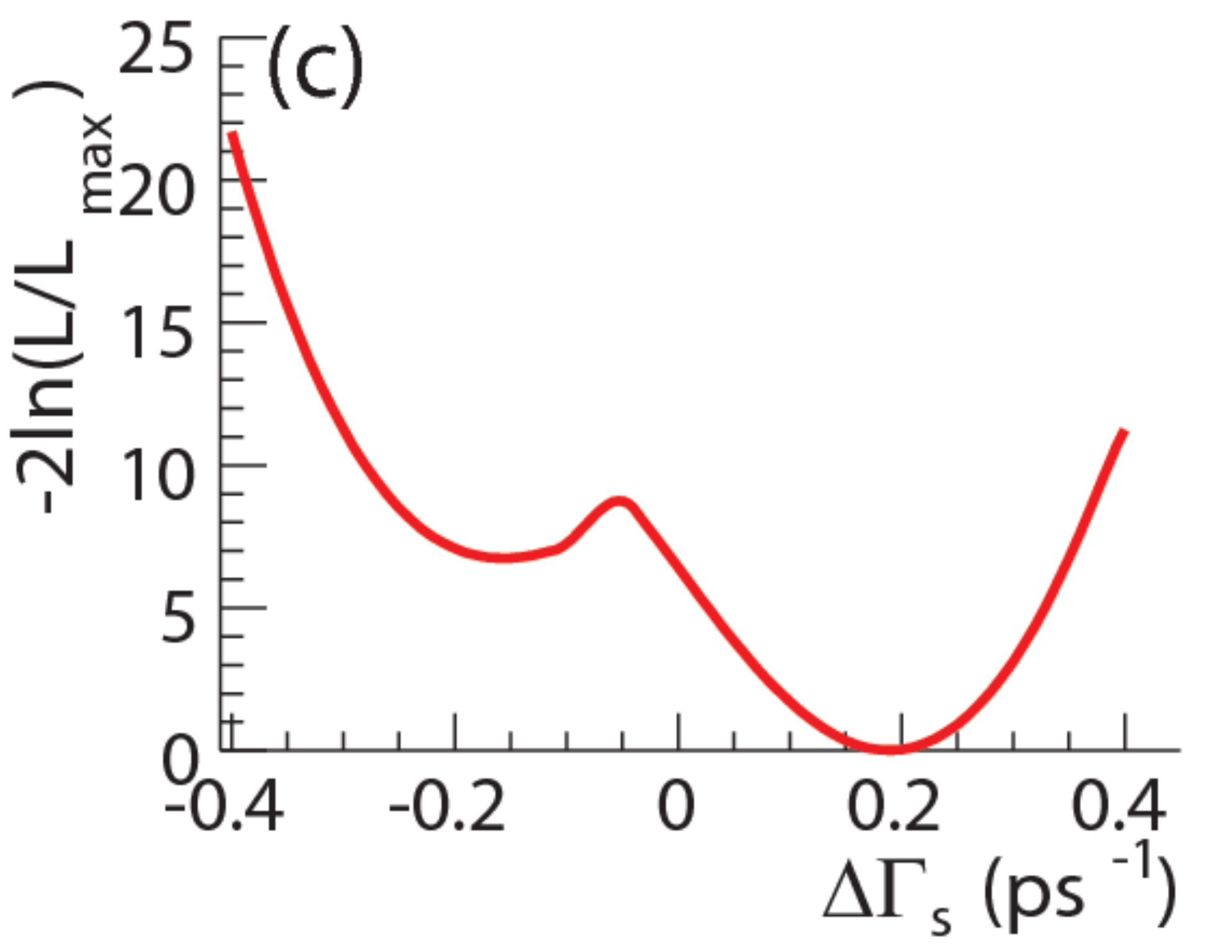}
\caption{D0's confidence regions in the $\Delta\Gamma_s-\phi_s^{J/\psi\phi}$ plane.  Shown are the 68.3\% C.L. 
(dashed) and 90\% C.L. contours, and the expectation from the standard model (black vertical line)
and mixing-induced $CP$ violation from new physics sources (green band). Below are one-dimensional 
likelihood profiles for $\phi_s^{J/\psi\phi}$ and $\Delta\Gamma_s$. }
\label{fig:eight}
\end{figure}

The D0 experiment employs a different strategy for dealing
with the twofold ambiguity in the tagged analysis.  They constrain
the strong phases $\delta_\parallel$ and $\delta_\perp$ to the world average
values\cite{ref:d0hfag} in the related decay $\bzero\rightarrow J/\psi K^{*0}$, within a Gaussian
uncertainty of $\pm \pi/5$.  Some recent theoretical work provides justification
for this approach: in reference~\cite{ref:RosnerGronau} the phases in the two systems are
estimated to be equal within ten degrees.  Results of the
constrained fits are shown in Table~\ref{table:d0Fits}.   Three
types of fit are performed:  a standard model fit, 
a $CP$ fit, and the new physics (NP) fit, using the previously discussed NP constraint.
The quantities ${\bar \tau}_s$, and the decay amplitudes are consistent with expectations
and with CDF's measurement.  D0 uses $\phi_s^{J/\psi\phi}$, the equivalent to $-2\beta_s$
in Eq.~\ref{eqn:finalAmp} as their $CP$ violation parameter\footnote{The nomenclature $\phi_s^{J/\psi\phi}$ has recently been invented
to designate this quantity, in order to distinguish it from the $\phi_s$ that we soon define, and from $-2\beta_s$ which by definition does not include new physics sources of mixing-induced $CP$ violation.}.  
Likelihood profiles in the space of ($\phi_s^{J/\psi\phi}, \Delta\Gamma_s$), as well as in $\phi_s^{J/\psi\phi}$ and $\Delta\Gamma_s$ separately, are
shown in Fig.~\ref{fig:eight}.  D0 gives point estimates 
$\phi_s^{J/\psi\phi}$ = -0.57 $^{+0.24}_{-0.30}$ (stat) $^{+0.07}_{-0.02}$ (syst) and 
$\Delta\Gamma_s$=0.19 $\pm$ 0.07 (stat) $^{+0.02}_{-0.01}$ (syst) ps$^{-1}$, based upon
their $CP$-fit. Using simulation,  D0 finds for the standard model 
a $p$-value of 6.6\%. The apparent discrepancy  (or fluctuation) goes in the same direction
as CDF.

\begin{table}[h]
\begin{center}
\caption{Results of fits to the $\bsToPsiPhi$ from the D0 collaboration.  In the first column is
the CP fit; in the second column is the SM fit, and in the third column is the NP fit. See the
text for further explanation.}
\begin{tabular}{|l|c|c|c|}
\hline \textbf{Parameter} & \textbf{D0 $CP$} & \textbf{D0 SM} & \textbf{D0 NP}   \\
\hline ${\bar \tau_s}$ (ps)                               & 1.52 $\pm$ 0.06         & 1.53 $\pm$ 0.06   & 1.49 $\pm$ 0.05\\
$\Delta\Gamma_s$ (ps)$^{-1}$                       & 0.19 $\pm$ 0.07         & 0.14 $\pm$ 0.07   & 0.083 $\pm$ 0.018\\
$A_\perp$                                       & 0.41 $\pm$ 0.04         & 0.44 $\pm$ 0.04   & 0.45 $\pm$ 0.03\\
$|A_0|^2 - |A_\parallel|^2 $                 & 0.34 $\pm$ 0.05         & 0.35 $\pm$ 0.04   & 0.33 $\pm$ 0.04\\
$\delta_1 = \delta_\perp - \delta_\parallel$       & -0.52 $\pm$ 0.42        & -0.48 $\pm$ 0.45  & -0.47 $\pm$ 0.42\\
$\delta_2 = \delta_\perp - \delta_0$               & 3.17 $\pm$ 0.39         & 3.19 $\pm$ 0.43   & 3.21 $\pm$ 0.40\\
$\phi_s^{J/\psi\phi}$                                           & -0.57$^{+0.24}_{-0.30}$ & $\equiv$ -0.04      & -0.46 $\pm$ 0.28\\
\hline
\end{tabular}
\label{table:d0Fits}
\end{center}
\end{table}

\section {Semileptonic Asymmetry}

Models with extra sources of mixing-induced $CP$ violation can have small amounts of $CP$ violation in 
the mixing (defined as $|q/p|\ne$1).  An observable quantity called the semileptonic asymmetry 
\begin{displaymath}
A^s_{SL} = 
\frac{d\Gamma/dt\left[\bsbar \rightarrow l^+X \right]-d\Gamma/dt\left[\bs \rightarrow l^-X \right]}
{d\Gamma/dt\left[\bsbar \rightarrow l^+X \right]+d\Gamma/dt\left[\bs \rightarrow l^-X \right]}
\end{displaymath}
related to $|q/p|$ through the definition
\begin{displaymath}
A^s_{SL} = \frac{1-|q/p|^4}{1+|q/p|^4}
\end{displaymath}
can be measured using semileptonic decays. In $\bs$ mesons a phase $\phi_s={\rm arg}(M_{12}/\Gamma_{12})$, where $M_{12}$
is a mass-matrix element and $\Gamma_{12}$ is a width matrix element, governs the size
of the asymmetry through the approximate relation (see Ref. \cite{ref:lenzNierste}) 
\begin{displaymath}
A^s_{SL}= \left |\frac{\Gamma^s_{12}}{M_{12}}\right |\sin{\phi_s}.
\end{displaymath}
The phase $\phi_s$ differs from $\phi_s^{J/\psi \phi}$ by a small shift, negligible compared to experimental resolution.  Since
\begin{displaymath} 
\left |\frac{\Gamma^s_{12}}{M^s_{12}} \right | = (49.7 \pm 9.4) \times10^{-4},
\end{displaymath}
this asymmetry can hardly be more than about half a percent.  Disregarding any theoretical input to $|\Gamma^s_{12}|$ while applying the relation $\Delta m_s = 2|M^s_{12}|$ and the 
NP constraint that $\Delta\Gamma_s = 2|\Gamma^s_{12}|\times \cos{\phi_s}$ one obtains
\begin{displaymath}
A^s_{SL}= \frac{\Delta\Gamma_s}{\Delta m_s} \tan{\phi_s}.  
\end{displaymath}
In constraining such models, then, one can choose as input either $|\Gamma^s_{12}|$, or
a measured value of $A^s_{SL}$, or both.  CDF,  using dimuon pairs in 1.6~fb$^{-1}$ of data, 
measures $A^s_{sl}$ = 0.020 $\pm$ 0.028\cite{ref:cdfasl} while D0, using both dimuon pairs and decays $\bs\rightarrow \mu\nu D_s$ with $D_s\rightarrow \phi \pi$ from 1.1~fb$^{-1}$ of data, measures $A^s_{SL}$ = 0.0001 $\pm$ 0.0090\cite{ref:d0asl}.  At this level of precision, the
theory value of 
$|\Gamma^s_{12}|$ is a more powerful constraint than the experimental value of $A^s_{SL}$, on new physics models. 
  
\section{Combined results}

\begin{figure*}[t]
\centering
\includegraphics[width=0.45\linewidth]{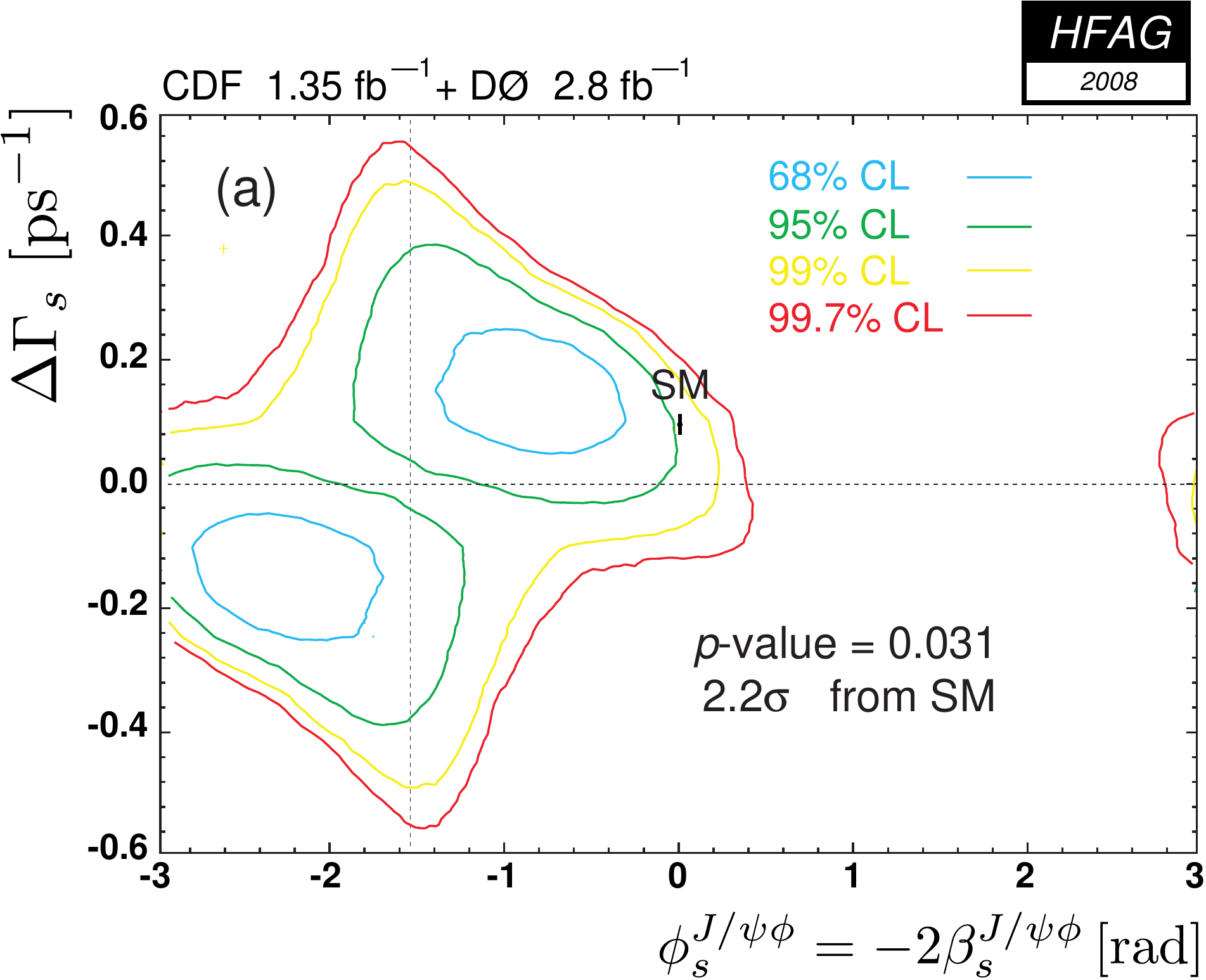}
\includegraphics[width=0.45\linewidth]{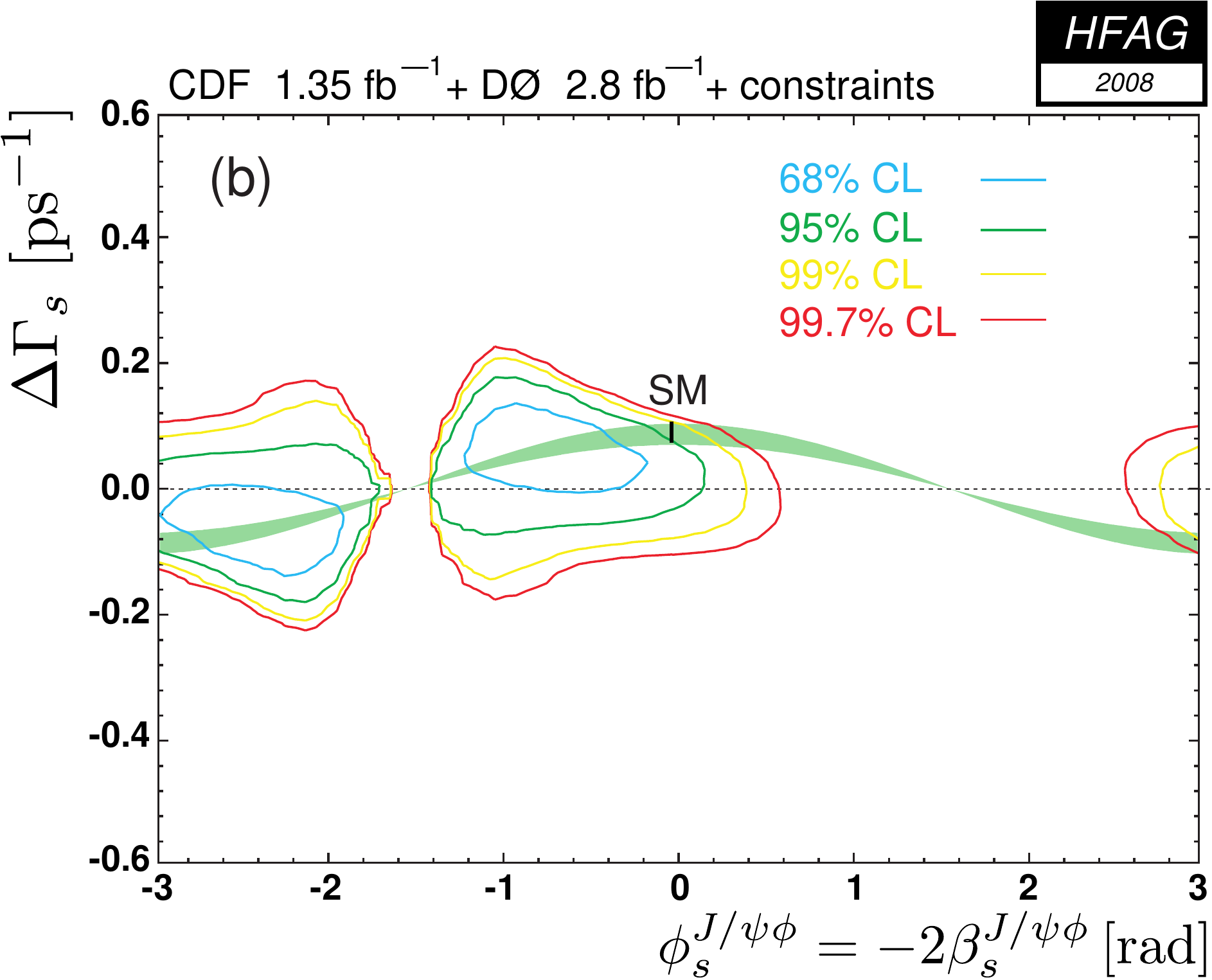}
\caption{HFAG Combinations of the D0 tagged analysis and the CDF tagged analysis. The standard
model is consistent with this data at the 2.2 $\sigma$ confidence level.  The plot on the 
right makes use of the semileptonic asymmetry $A^s_{SL}$}\label{fig:hfag}
\end{figure*}

The analyses of $\bsToPsiPhi$ from CDF and D0 are compatible with each other,
and with the the standard model at only the 15\% C.L. (CDF) and the 6.6\%
C.L.(D0). Following a new analysis of the D0 data, in which the strong phase constraints
were dropped, HFAG has combined the two analyses.  Details
of the procedure can be found in Ref.~\cite{ref:Combo}.  The combined
contours are shown in Fig.~\ref{fig:hfag}.  The $p$-value for the
combined result is 3.1\%, corresponding to 2.2 Gaussian standard
deviations.

\section {Conclusion}

Discrepancy?  Or fluctuation?  Today, the only known source of $CP$
violation in the physics of elementary particles is the CKM mechanism,
arising from the Higgs-Yukawa sector of the three-generation standard
model. A firmly established discrepancy between the predicted value of
$\beta_s$ and the standard model value would imply new sources of $CP$
violation, possibly from heavier particles out of the reach of today's
accelerators.  It could have a broader impact as well, and shed light
on the baryon asymmetry of the universe.  Unfortunately, with the
present uncertainty (statistical, mostly) and at the present
significance (2.2~$\sigma$) the question is not yet settled.

\begin{acknowledgments}
The author thanks the conference organizers for
a very enjoyable week in Melbourne. 
\end{acknowledgments}

\bigskip 

\end{document}